\documentclass[aps, prl, twocolumn, 10pt]{revtex4-1}
\usepackage[utf8]{inputenc}
\usepackage{amsmath, amssymb}
\usepackage{graphicx}
\usepackage{bm}
\usepackage{tipa} 
\newcommand{\thorn}{\mbox{\textthorn}} 

\begin{document}

\title{The Universe in four letters: a bedtime history}

\author{Pedro Bargueño}
\affiliation{Departamento de Física, Universidad de Alicante, E-03690 Alicante, Spain}

\begin{abstract}
One evening, a father tells his three daughters, Lucía, Inés, and Ana, a bedtime story unlike any other. It is a tale of four silent architects ($c$, $\hbar$, $G$, and $\Lambda$) whose presence or absence shapes the very fabric of reality. By imagining worlds where each constant is removed, they explore realms from the quantum gravity of the Planck scale to the stretching cosmos of de Sitter space, from the mass of the observable Universe to the surprising link with the proton's whisper. Woven with the insights of great scientists, this narrative reveals how these fundamental constants write the story of everything, from the smallest particle to the vast cosmic horizon, suggesting a beautiful and hidden unity in the architecture of our world.
\end{abstract}

\maketitle

\begin{center}
{\it For Lucía, Inés and Ana}

\end{center}

\section*{The Pillars of the World}

In the small village of Baides, the fire crackled softly as three sisters (Lucía, who gazed at distant galaxies; Inés, who wondered about the tiniest particles; and Ana, who sought the patterns connecting everything) curled up on the rug. Their father closed his book, full of equations and marvelous symbols, such as $\thorn$ and  $\eth$, and smiled.
\\
\\
“A true story?” he asked. “Then I will tell you the truest one I know. It is the tale of the four silent architects of our Universe. They are not characters, but constants: the speed of light $c$, Planck's constant $\hbar$, Newton's constant $G$, and the cosmological constant $\Lambda$. Like invisible rulers, they decree how long a second is, how heavy a gram feels, and how vast space can be \cite{barrow2002}.”
\\
\\
“But Papa,” Ana interrupted, reading over his shoulder, “this note says the cosmological constant might not be truly constant forever \cite{lambdafootnote}.”
\\
\\
“A very sharp observation, Ana,” he nodded. “In modern cosmology, $\Lambda$ is likely an effective scale that can evolve, like dark energy. But for our story tonight, we'll treat it as a fixed landmark, a guidepost in the vastness. The profound idea that all physical scales stem from a few numbers was championed by thinkers like Eddington \cite{eddington1923}, Dirac \cite{dirac1937,dirac1938}, and Zeldovich \cite{zeldovich2008}. As Barrow and Tipler explored, these constants make our particular Universe possible \cite{barrowtipler1986}. But the most elegant analysis of how they create ‘cosmic coincidences’ came from Carr and Rees \cite{carrrees1979}. Tonight, I will follow their spirit.”
\\
\\
“How?” asked Lucía.
\\
\\
“By a game of imagination,” he said. “We will consider our set $\{c, \hbar, G, \Lambda\}$ as defining a four-dimensional space of all possible worlds. Then, we'll visit the worlds that appear when we formally remove one constant at a time. Each world is a coherent realm with its own rulers of length, time, and mass.”

\section*{Realm I: The Quantum Gravity Domain ($\Lambda \to 0$)}

“Our first journey,” he began, “is to a world where the cosmological constant $\Lambda$ fades away. The remaining trio—$c$, $\hbar$, and $G$—build the most famous of scales: the Planck units \cite{planck1899}.”

\[
l_p = \sqrt{\frac{G\hbar}{c^3}}, \quad t_p = \sqrt{\frac{G\hbar}{c^5}}, \quad m_p = \sqrt{\frac{\hbar c}{G}}.
\]
\\
\\
“This, Lucía, is the kingdom of the unimaginably small,” he said. “The Planck length $l_p$ is where the ideas of ‘here’ and ‘there’ dissolve. To probe a smaller distance would require so much energy you’d create a black hole in your experiment. It is the regime where quantum mechanics, relativity, and gravity become inseparable. Here, the Compton wavelength of a particle, $\hbar/(mc)$, equals its Schwarzschild radius, $2Gm/c^2$, exactly at the Planck mass $m_p$. This is the domain of the very early Universe and black hole thermodynamics, where Hawking showed black holes could radiate \cite{hawking1975}.”

\section*{Realm II: The Quantum de Sitter Universe ($G \to 0$)}

“Now,” he continued, “let us visit a stranger world. Let Newton's constant $G$ vanish. We keep $\hbar$, $c$, and $\Lambda$. Without dynamic gravity, what rules this world?”
\\
\\
“Nothing pulls on anything?” Inés asked.
\\
\\
“Correct. But $\Lambda$ remains, giving empty space a constant curvature. The symmetry of this Universe is de Sitter invariance \cite{gibbons1977}. Its natural scales come from $\Lambda$ alone:”
\[
l_{\mathrm{dS}} = \Lambda^{-1/2}, \quad t_{\mathrm{dS}} = c^{-1} \Lambda^{-1/2}.
\]
“These are the Hubble radius and the age of our \textit{actual} observable Universe. But the most fascinating ruler is the mass:”
\[
m_{\mathrm{dS}} = \frac{\hbar}{c} \Lambda^{1/2}.
\]
\\
“This mass is incredibly tiny. Its Compton wavelength equals the de Sitter horizon radius $l_{\mathrm{dS}}$. It is the lightest quantum particle that can fit within this stretching cosmos. This scale also appears as the Gibbons–Hawking temperature of empty space, $k_B T_{\mathrm{dS}} \sim \hbar c \sqrt{\Lambda}$ \cite{gibbons1977}.”
\\
\\
He looked at their curious faces. “And geometry alone, without $G$, restricts what can exist. Higuchi showed a massive spin-2 field must obey $m^2 \ge (2/3)(\hbar^2/c^2)\Lambda$ to be consistent here \cite{higuchi1987}. So $\Lambda$ acts as a cosmic filter.”

\section*{Realm III: The Classical Cosmological Realm ($\hbar \to 0$)}

“For our third journey,” said the father, “we let the quantum constant $\hbar$ fade to zero. We keep $\Lambda$, $c$, and $G$. This is the world of pure, classical cosmology.”
\\
\\
“No quantum fuzziness?” Inés asked, slightly disappointed.
\\
\\
“None. The length and time are still from $\Lambda$: $l_o = \Lambda^{-1/2}$, $t_o = c^{-1}\Lambda^{-1/2}$. But the mass ruler changes because $G$ is back:”
\[
m_o = \frac{c^2}{G} \Lambda^{-1/2}.
\]
\\
“Can you guess its value?” he asked Lucía. “About $10^{53}$ kg. This is the estimated total mass within our \textit{observable Universe} \cite{peebles1993}! It's no coincidence. $\Lambda$ and $G$ give a critical density $\rho_c = \Lambda c^2/(8\pi G)$. The mass inside a sphere of radius $\Lambda^{-1/2}$ is of order $m_o$. So in this classical limit, $\Lambda$ sets the scale for all the stuff in the cosmos.”
\\
\\
He leaned forward. “There's another clue. The acceleration scale $a_0$ in Milgrom's Modified Newtonian Dynamics (MOND), proposed to explain galaxy rotation without dark matter, is numerically close to $c^2\sqrt{\Lambda}$ \cite{milgrom1983a,milgrom1983b,milgrom1983c}. Perhaps the edge of Newtonian dynamics is a signal from the global, $\Lambda$-dominated geometry.”

\section*{Realm IV: The Hadronic Whisper ($c \to \infty$)}

“Our final world is the most subtle,” the father said, his voice low. “What if the speed of light $c$ became infinite? This is the non-relativistic limit. The symmetry becomes the Newton-Hooke group. The length and time are still from $\Lambda$, but the mass ruler becomes:”

\[
m_{\mathrm{nr}} \sim \left(\frac{\hbar^2\sqrt{\Lambda}}{G}\right)^{1/3}.
\]
\\
\\
“Calculate this,” he said, “and you find a mass around $10^{-27}$ kg. Inés, what particles live here?”
\\
\\
“Protons! Pions!” she exclaimed.
\\
\\
“Exactly,” he smiled. “Weinberg noted this as a curious dimensional coincidence \cite{weinberg1972}. But here, it emerges as the mass scale for a 
non-relativistic quantum world in a $\Lambda$-cosmos. It hints that $\Lambda$, the constant of the vast cosmos, might be linked to the mass of matter's building blocks.”

\section*{An Ancient and Charged Epilogue}

Ana, who had been quiet, finally spoke. “Papa, you told a tale of four, but the manuscript also talks about electromagnetism and a charge $e$.”
\\
\\
“Ah, the splendid epilogue,” he said. “Long before Planck, Stoney tried to unify gravity and electromagnetism \cite{stoney1881}. Using $e$, $G$, and $c$ (and setting $\hbar=0$, $\Lambda=0$), he defined Stoney units:”

\[
l_s = \sqrt{\frac{G e^2}{4 \pi \epsilon_0 c^4}}, \quad m_s = \sqrt{\frac{e^2}{4 \pi \epsilon_0 G}} , \quad  t_s = \sqrt{\frac{G e^2}{4 \pi \epsilon_{0} c^6}}
\]
\\
\\
“The Stoney mass $m_s$ is the mass of an extremal charged black hole. It's a purely classical unification. Planck’s genius was to include $\hbar$, refining this into quantum gravity scales. Indeed, $l_s \sim \sqrt{\alpha} l_p$, where $\alpha = e^2/(4\pi\epsilon_0\hbar c) \approx 1/137$ is the fine-structure constant.”
\\
\\
“This leads to a beautiful geometric idea,” he said, drawing a circle. “For electromagnetism, $\alpha$ measures coupling strength. For gravity, we can ask: how close is a quantum object to becoming a black hole? Take an object of mass $m$. Its quantum size is its Compton wavelength $R \sim \hbar/(m c)$. The Hoop Conjecture by Thorne \cite{thorne1972} says an object forms a black hole if compressed within a circumference $C \lesssim 4\pi G m/c^2$.”
\\
\\
“So,” Ana interrupted, “the ratio of its gravitational radius to its quantum size...”
“Gives a dimensionless number: $\Theta \sim \frac{G m}{R c^2} \sim \frac{G m^2}{4\pi \hbar c}$. This is the compactness of a quantum object. If $\Theta \ll 1$, you have a normal particle. If $\Theta \sim 1$, you are at the Planck mass $m_p$. If $\Theta \gg 1$, collapse is inevitable. It's a single number linking quantum, gravity, and geometry.”

\section*{The Moral of the Tale}

The fire had died to embers. Lucía, Inés, and Ana were quiet, their minds swimming with realms of scale.
\\
\\
“So the moral, my daughters,” their father concluded, “is not that constants can vanish, but that each organizes reality uniquely. Their presence or absence defines possible worlds. The astounding fact, as Carr and Rees articulated so well \cite{carrrees1979}, is that the scales they build—from Planck to Hubble, from Universe mass to proton mass—are not random. They are the architecture of our one, real, beautiful Universe.”
\\
\\
“The cosmological constant $\Lambda$, even if not perfectly constant \cite{lambdafootnote}, is not a passive number. It actively shapes both cosmic expanse and, perhaps, the mass of matter within it. This tale of four constants suggests the very large and very small are woven together by a deeper, yet undiscovered, principle.”
\\
\\
He kissed each on the forehead.
“And that is a story for another night.”
\\
\\
P. B. acknowledges financial support from the Generalitat Valenciana through PROMETEO PROJECT CIPROM/2022/13.


\begin{thebibliography}{99}
\bibitem{barrow2002} J. D. Barrow, \textit{The Constants of Nature: from Alpha to Omega} (Vintage, London, 2002).
\bibitem{lambdafootnote} M. Abdul-Karim {\it et al.} (DESI Collaboration), Phys. Rev. D {\bf 112}, 083515 (2025); see also references therein.
\bibitem{eddington1923} A. S. Eddington, \textit{The Mathematical Theory of Relativity} (Cambridge University Press, Cambridge, 1923).
\bibitem{dirac1937} P. A. M. Dirac, Nature \textbf{139}, 323 (1937).
\bibitem{dirac1938} P. A. M. Dirac, Proc. R. Soc. A \textbf{165}, 198 (1938).
\bibitem{zeldovich2008} V. Sahni and A. Krasinski, Gen. Relativ. Gravit. \textbf{40}, 1557 (2008).
\bibitem{barrowtipler1986} J. D. Barrow and F. J. Tipler, \textit{The Anthropic Cosmological Principle} (Oxford University Press, Oxford, 1986).
\bibitem{carrrees1979} B. J. Carr and M. J. Rees, Nature \textbf{278}, 605 (1979).
\bibitem{planck1899} M. Planck, Sitzungsber. Preuss. Akad. Wiss. \textbf{5}, 440 (1899).
\bibitem{hawking1975} S. W. Hawking, Commun. Math. Phys. \textbf{43}, 199 (1975).
\bibitem{gibbons1977} G. W. Gibbons and S. W. Hawking, Phys. Rev. D \textbf{15}, 2738 (1977).
\bibitem{higuchi1987} A. Higuchi, Nucl. Phys. B \textbf{282}, 397 (1987).
\bibitem{peebles1993} P. J. E. Peebles, \textit{Principles of Physical Cosmology} (Princeton University Press, Princeton, 1993).
\bibitem{milgrom1983a} M. Milgrom, Astrophys. J. \textbf{270}, 365 (1983).
\bibitem{milgrom1983b} M. Milgrom, Astrophys. J. \textbf{270}, 371 (1983).
\bibitem{milgrom1983c} M. Milgrom, Astrophys. J. \textbf{270}, 384 (1983).
\bibitem{weinberg1972} S. Weinberg, \textit{Gravitation and Cosmology} (Wiley, New York, 1972).
\bibitem{stoney1881} G. J. Stoney, Philos. Mag. \textbf{11}, 381 (1881).
\bibitem{thorne1972} K. S. Thorne, in \textit{Magic Without Magic}, ed. J. R. Klauder (Freeman, San Francisco, 1972).
\end{thebibliography}
\end{document}